# Studies of Ni-Cr complexation in FLiBe molten salt using machine learning interatomic potentials

*Siamak Attarian [a*], Dane Morgan [a*], Izabela Szlufarska [a*]*

[a] Department of Materials Science and Engineering, University of Wisconsin, 1509 University Ave, Madison, WI, 53706, USA

* sattarian@wisc.edu

* ddmorgan@wisc.edu

* szlufarska@wisc.edu

**Abstract:**

In nuclear and/or solar applications that involve molten salts, impurities frequently enter the salt as either fission products or via corrosion. Impurities can interact and make complexes, but the impact of such complexation on the properties of the salts and corrosion rates has not been understood. Common impurities in molten salts, such as FLiBe, include Cr, Ni, and Fe. Here, we investigate the complexation of Cr and Ni in FLiBe using molecular dynamics based on a machine learning interatomic potential (MLIP) fitted using the atomic cluster expansion (ACE) method. The MLIP allows us to overcome the challenges of simultaneously needing accurate energetics and long time scale to study complexation. We demonstrate that impurity behavior is more difficult to capture than that of concentrated elements with MLIPs due to less sampling in training data, but that this can be overcome by using active learning strategies to obtain a robust fit. Our findings suggest that there is a weak but potentially significant binding free energy between $CrF_2$ and $NiF_2$ in eutectic FLiBe of $-0.112\ eV$. Under certain conditions this binding creates a significant short-range order between the species and lowers the redox potential of $NiF_2$ in the presence of $CrF_2$ in FLiBe, making Ni dissolution more favorable in the presence of Cr as compared to its dissolution in pure FLiBe. However, we find little impact of this complexation on the diffusivity of Ni and Cr. Overall, the methodology presented here suggests an approach to modeling complexation with MLIPs and suggests that interactions between dissolved cations could be playing a significant role in some salt thermophysical properties.

## 1. Introduction

Molten salts are being considered as coolants and fuel salts in nuclear reactors due to their excellent thermal conductivity and as storage medium in solar power plants due to their high heat capacity [1]. In this work, we focus on eutectic FLiBe (66.6% LiF – 33.3% $BeF_2$), which is a candidate salt for the next generation molten salt reactors (MSR), and its thermophysical properties and corrosivity are being actively investigated by researchers through experimental and theoretical means [2–6]. Impurities in FLiBe can enter the salt via corrosion of the structural materials and can affect the properties of the salt such as density, diffusivity, and corrosivity. The typical structural materials that are used in reactors and exposed to molten salts are metal alloys and can contain a range of alloying elements (e.g., Cr, Ni, Fe, W, Co). The electrochemical potential of the salt is controlled by the exact composition of the FLiBe and its impurity content. The salt potential has a lower bound set by the stability of $BeF_2$ at $T$ = 973 K at about -4.7 V, and various measurements of the salt potential at different points of the purification show a range of -4.19 V to -2.86 V, as referenced to $F_2/F^-$ [7]. For most of this range, Ni is quite stable ($NiF_2$ redox potential is -3.05 V [8]), but Cr is quite soluble ($CrF_2$ redox potential is -3.95 V). As most structural metal alloys contain some Cr, its instability in a solid phase is a concern. This concern has led to a focus on Ni-based alloys for FLiBe exposed structural materials (e.g., Hastelloy), which can form a protective Ni-rich layer. The essential roles of Cr and Ni in relevant alloys for salt exposed materials suggest that they may often coexist in the salt and therefore undergo potential interaction.

To our knowledge there is no study to date exploring the strength of this interaction and if it might produce Cr-Ni complexes that impact the salt properties or the behavior of Ni and Cr in the salt, e.g., their diffusivities, corrosion potentials, or corrosion rates.

Molecular dynamics (MD) simulations provide a powerful tool to explore Cr-Ni interactions in FLiBe but there are certain challenges. Specifically, correctly simulating the coupling may require (i) highly accurate energetics and capturing of subtle electronic effects associated with the effects of bringing Cr and Ni together in the molten salt and (ii) long MD simulations of tens of nanoseconds to obtain well-converged properties of the system at various inter-solute distances between Ni and Cr. These requirements cannot be easily met with Ab Initio MD (AIMD) due to the long time scales required by (ii) or with classical potentials such as Polarizable Ion Model [9] (PIM), Born-Mayer-Huggins (BMH) [10], etc., due to the delicate balance of energetics required by (i). In this work, we demonstrate that MLIPs are able to achieve both the desired accuracy and time scales and we use these potentials to determine Ni-Cr coupling in molten salts.

MLIPs have been successfully used to model a number of molten salts in the past few years [2,3,11]. Generally, to fit an MLIP one needs to generate several atomic configurations and obtain their energies and forces using density functional theory (DFT) calculations. Each atomic configuration typically consists of a supercell of about 50-200 atoms. The energy and atomic forces of each atomic configuration constitute one training configuration or one training data. A successful potential fitting requires a significant amount of training data (typically 1000 to 10000 different calculations of one cell's energy and forces) to provide a sufficient number of atomic environments for a robust fit. Fitting an MLIP for dilute impurity species, which is our focus in this work, is more challenging compared to pure salts since every training data point (a DFT calculation of one set of atomic positions) contains only a few or even just one atom of the dilute species, and consequently one atomic environment around that species. This limits the amount of information related to the dilute species that is being provided by each training data. The limited data necessitates a more elaborate way of generating the training data compared to the case of fitting a potential for a salt with a larger mole fraction of each species, e.g., a typical two-component or eutectic salt. To the extent of the authors' knowledge, no prior work has been published on fitting MLIPs for dilute impurities in molten salts.

Here we use the ACE potential [12] to fit parameters for FLiBe with Cr and Ni impurities and study their binding energy by calculating the potential of mean force (PMF) between them. We provide a strategy based on the available active learning methods in the literature to generate sufficient training data. Then we run multiple MD simulations and study the effects of proximity of Cr and Ni in FLiBe. We find that there is a weak binding between $CrF_2$ and $NiF_2$, which causes formation of short-range order between the species that may affect properties such as diffusivity, viscosity, thermal conductivity, etc. However, we show that the weak binding does not significantly affect the diffusivity of Cr and Ni. This binding also lowers the redox potential of $NiF_2$ in the presence of $CrF_2$ in FLiBe, making Ni dissolution more favorable in the presence of Cr compared to dissolution of Ni in pure FLiBe. In Section 2 we discuss our methodology including the simulation details, formulation of ACE, data generation, and fitting strategies. In section 3 we present and discuss the results and in section 4 we give the conclusions.

## 2. Methodology

### 2.1. Simulation details

The free energy of binding between two solutes can be obtained by calculating their PMF. The most straightforward approach to calculate PMF (or W(r)) between two particles is via their radial distribution function (RDF or g(r)) using the following equation [13,14]:

$$W(r) = -k_B T ln\big(g(r)\big) \qquad\qquad Eq.\,1$$

where $k_B$ is the Boltzmann constant and $T$ is the temperature. Obtaining PMF from Equation 1 for dilute systems requires a well-converged g(r) between the solutes which in turn requires a long MD simulation for adequate sampling at various inter-solute distances [15–17]. While some researchers choose different techniques such as thermodynamic integration (TI) [18], umbrella sampling (US) [19], etc., where they perform several shorter simulations to calculate PMF, it has been shown that the PMFs calculated from g(r) that is obtained from even quite practical simulations (8 ns in Ref. [20]) or from systems with higher solute concentrations [21], are consistent with the results of techniques such as TI and US. We will use W(r) to calculate the binding free energy for Ni and Cr as

$$\Delta G_b = \min\big(W(r)\big) - W_{infinite}(r) \qquad\qquad Eq.\,2$$

where $r$ is the distance between the Ni and Cr and W(r) is calculated from the Ni-Cr pair correlation function. This $\Delta G$ is the change in free energy of the systems between when the Ni-Cr are approximately infinitely far apart and when they are at their most stable bound separation. This free energy includes all contributions from atomic motion except that from the Ni and Cr, which are considered fixed. It therefore does not include the Ni and Cr configurational entropies.

In this work, we used the PMF obtained from Equation 1 to calculate the binding energy between Cr and Ni solutes. Modeling dilute systems requires a large enough simulation cell to avoid unwanted image interactions and this can put high computational demands on the simulation time needed to converge g(r). To be sure our results are converged with respect to size we ran three separate simulations using systems that contain 202 atoms (FLiBeCrNi_small system), 398 atoms (FLiBeCrNi_medium system) and 594 atoms (FLiBeCrNi_large system) and compared the g(r) between Cr-Ni in each system. These systems had 28, 56, and 84 units of $Li_2BeF_4$ respectively, plus 1 unit of $CrF_2$ and 1 unit of $NiF_2$. See Figure 1 for more details.

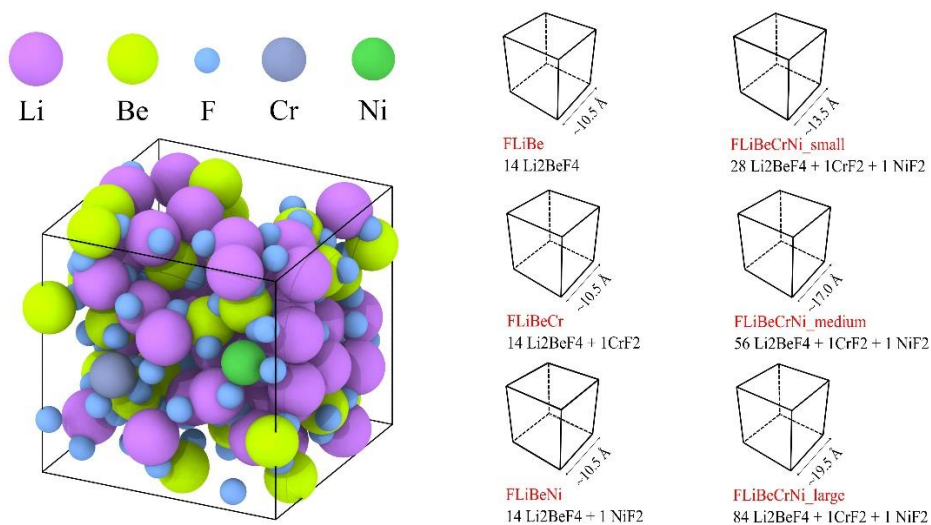

**Figure 1.** *A typical simulation cell containing Li, Be, F, Cr, and Ni (left), and the systems that were used in this work either as the training data or for MD simulations (right). The unit cell length shown in the graph for each cell is an average value.*

The MD simulations were performed using the LAMMPS [22] software package under the constant pressure – constant temperature (NPT) ensemble at $T = 973$ K and $P = 0$ bar for 200 ns. After running the MD simulations, we calculated the g(r)s from the cumulative data up to each time $\tau$, evaluating for $\tau = 20, 40, \ldots 200$ ns. As all these simulations are considered at equilibrium the convergence of this cumulative average allows us to assess how much time is needed to converge the g(r) and associated W(r). We then calculated the Warren-Cowley parameter for the relative position of Cr and Ni in the FLiBeCrNi_large system to assess the extent of association of two impurities. We also performed a separate set of simulations to compare the diffusivity of each impurity when they are isolated and when they are combined to determine the effects of the complexation on the kinetics of the impurities. For the isolated cases, we used two simulation cells, one with 14 units of $Li_2BeF_4$ and 1 unit of $CrF_2$ (FLiBeCr system containing 101 atoms) and another with 14 units of $Li_2BeF_4$ and 1 unit of $NiF_2$ (FLiBeNi system containing 101 atoms. See Figure 1). For the combined case we used the FLiBeCrNi_small system. The ACE potential was used to model the interaction between the Li, Be, F, Cr, and Ni ions. The details of the ACE potential and the fitting methodology are described in the following sections. We only consider the complexation of a single Cr and a single Ni, as we assume this will be the dominant coupling and to keep the scope of work in the paper practical. More complex complexations, e.g., Ni with 2 Cr, might be of interest for future studies.

## 2.2. ACE interatomic potential

The atomic cluster expansion or ACE was introduced in 2019 [12]. Like many other MLIPs [23–25], ACE divides the energy of each atomic configuration into the sum of the energy contributions of individual atoms. The energy contribution of each atom is then divided into the sum of its two-body, three-body, etc., interactions with the atoms within a certain cut-off distance. ACE has a linear formulation as follows

$$\varphi_i = \sum_v \tilde{c}_v \boldsymbol{A_{iv}} \qquad\qquad Eq.\,3$$

where $\varphi_i$ is the energy contribution of atom $i$, $\tilde{c}_v$s are the fitting parameters, and $\boldsymbol{A_{iv}}$s are the many-body basis functions of body order $v + 1$, calculated by the following equation

$$\boldsymbol{A_{iv}} = \prod_{t=1}^{v} A_{iv_t} \qquad\qquad Eq.4$$

The atomic bases $A_{iv}$ are obtained by projecting single-bond basis functions $\phi_v$ on the atomic density $\rho_i$

$$A_{iv} = \langle \phi_v | \rho_i \rangle \qquad\qquad Eq.5$$

where

$$\rho_i = \sum_{j}^{j \neq i} \delta_{\mu_i \mu_j} \delta(r - r_{ji}) \qquad\qquad Eq.6$$

and

$$\phi_{\mu_i \mu_j nlm} = R_{nl}^{\mu_i \mu_j}(r_{ji}) Y_{lm}(\hat{r}_{ij}) \qquad\qquad Eq.7$$

In the above equations, $\mu$'s are the atomic species and $r_{ji}$ is the displacement vector from $i$ to $j$. $\phi$'s are taken as irreducible basis functions of the rotation group, meaning a radial function $R$ multiplied by a spherical harmonics $Y_{lm}$. The choice of the radial function is arbitrary, and, in our work, we used Chebyshev polynomials. Detailed discussion about the ACE formulation can be found in references [12,26]. To fit the ACE potential, we used the pacemaker code [26] developed by the authors of ACE and to run MD simulations we used the pace pair style in LAMMPS.

## 2.3 Training data generation with active learning

During the simulation of the FLiBeCrNi_large system, Cr and Ni may get close to each other or get separated to the extent that would be considered isolated ions. To help assure that the ACE potential to work for all the possible scenarios that arise in the simulations, we used 4 series of training data: configurations that only included $Li_2BeF_4$ units (FLiBe system), configurations that only contained one of the solutes (FLiBeCr and FLiBeNi systems as described in Section 2.1 and Figure 1), and configurations that contained both solutes (FLiBeCrNi_small system). We collected all the required configurations and fitted one ACE potential. There are many approaches to generate training data. One popular approach is to run several AIMD simulations at different densities and temperatures for each system and use a portion of generated atomic configurations as training data. This method has successfully been adopted in many studies[27,28], but it has been shown that it is not the most efficient approach for data generation [2]. It is especially important to be efficient in generating training data for systems with impurities as there is only one atom of that species in each simulation cell and only limited data about that atom (atomic force of the impurity, energy contribution of the impurity to the whole system) is being provided to the ACE potential during the fitting. Therefore, simply running several AIMD simulations may not provide enough atomic environments around the impurity atom unless the AIMD simulations are much longer compared to the case of fitting a potential for pure FLiBe, which is not computationally desirable. It is possible that fitting to more concentrated Ni and Cr in the salt would have provided better sampling and given accurate results transferable to the dilute cases. However, these higher concentrations would introduce more interactions between Ni with itself and Cr with itself, potentially requiring more fitting parameters and training data. We did not explore this approach but determining the most efficient method to model dilute impurities is an important topic for further study.

For training data generation, we used an active learning scheme based on the D-optimality criterion [29,30]. Generally, active learning or active sampling means collecting the minimum required training data for fitting a machine learning model. In the context of molecular dynamics, active learning methods include generating a huge number of atomic configurations that have not been DFT calculated, and then selecting a minimum set of them, either all at once or iteratively, that contain the most diverse atomic environments. The selected atomic configurations are then DFT calculated, and their energies and forces are used as the training data. The active learning procedure that we followed in this study included, 1- generating initial training data using AIMD simulations, 2- fitting an ACE potential, 3- running MD simulations with the fitted potential to generate more atomic configurations, 4- selecting new atomic configurations using the D-optimality criterion, 5- DFT calculation of the selected atomic configurations and adding them to the training data, and then repeating steps 2 to 5 a few times until sufficient training data are collected as discussed in the following. This procedure is shown schematically in Figure 2.

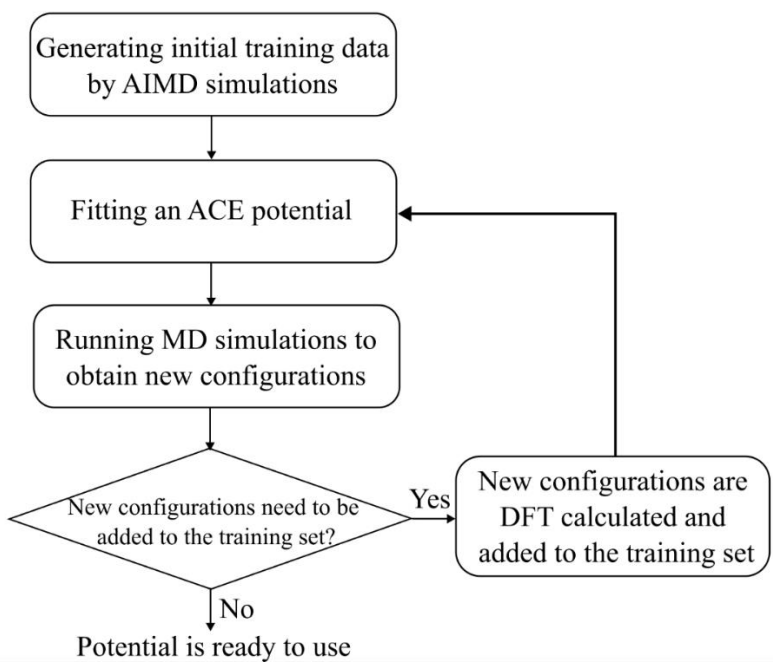

*Figure 2*. *The active learning procedure followed in this study to fit ACE potentials.*

To describe the D-optimality criterion briefly, consider that the potential featurizes each atom by N features. From the original training data, it is possible to collect $N$ atoms that would create an $N \times N$ matrix with the highest modulus of the determinant. This set of atoms is called the active set. Now every time a new configuration is being assessed, its atoms are featurized and replaced with each row of the active set one by one. If in any of these replacements, the modulus of the determinant is increased, that atomic configuration will be added to the training set (after DFT calculation) and the active set will be updated. Detailed information about this method is described in [30].

To generate training data for each of the aforementioned four systems, we started by running 7 AIMD simulations each for 0.5 ps with a time step of 1 fs at temperature 1100 K and densities between 1.7 gr/cm$^3$ < $\rho$ < 2.2 gr/cm$^3$ in constant volume – constant temperature (NVT) ensemble. We selected every 10$^{th}$ timestep of these simulations (overall 350 atomic configurations for each system) and used them as the initial training data and fitted an ACE potential. Later we used the fitted potential to run several MD simulations at temperatures between 600 K < $T$ < 1600 K and pressures between -1 GPa < $P$ < 10 GPa to generate more atomic configurations and select them based on the active learning procedure. After the DFT calculation of the selected data, we retrained the potential and repeated the aforementioned procedure with the new active set and newly trained potential. For the FLiBe system, initially, we used FLiBe cells with 98 atoms (28 Li, 14 Be, and 56 F), and at the later stages of the training we also ran MD simulations of FLiBe cells with 196 atoms to include more diverse training data. We stopped the active sampling procedure when during a 1 ns MD simulation less than 1% of the atomic configurations were qualified to be added to the training data. To generate testing data, we ran MD simulations for 1 ns at the pressure and temperature range mentioned above and randomly selected 500 configurations for each system.

We used a cutoff radius of 10 Å [2]. The hyperparameters of the potentials (the number of many-body orders, radial functions, and spherical harmonics) were determined by starting from lower numbers and increasing the number of each hyperparameter until there was no meaningful increase in the accuracy of the potential (fitting errors) by increasing the number of parameters.

DFT+U calculations were performed using the VASP 6.3.0 package [41] and by considering spin polarization. PBE-GGA approximation [43] was used for the exchange-correlation functional and the D3 method of Grimme [37] was used to account for dispersion forces. PAW-PBE potentials which were used in this study are Li_sv ($1s^2 2s^1$), Be($2s^2$), F ($2s^2 2p^5$), Cr_pv ($3p^6 3d^5 4s^1$) and Ni_pv ($3p^6 3d^8 4s^2$). An energy cutoff of 600 eV was used for the plane-wave basis set and a single gamma point was used to sample the Brillouin zone. The calibrated Hubbard values that were used in this work are U = 3.7 and L = 0 for Cr and U = 6.2 and L = 0 for Ni, which are obtained from the materials project website [31]. These values are benchmarked for solid oxides so their application in the present work could lead to some errors. The simplified rotationally invariant approach of DFT+U was used [32]. We ran several AIMD simulations of Cr in FLiBe without fixing the magnetic moment and in most of the simulations the magnetic state was 4. However, at some timesteps, the spin would flip to 2, stay at 2 for a short while, and then flip back to 4. By taking out the configurations at those timesteps and running single point DFT calculations at magnetic states of 2 and 4, we realized that the magnetic state of 4 has lower energies compared to the magnetic state of 2 for most configurations. It therefore seems the predicted spin of 2 is usually due to the numerical optimization converging to a metastable spin state. It should be noted that the calculated atomic forces for Cr were noticeably different between calculations with spins 2 and 4 so we couldn't mix the data and fit the potential for a single Cr atom type. Hence, in all our DFT calculations, we fixed the magnetic moment of chromium on 4. For nickel as $NiF_2$ in FliBe, AIMD with unfixed spin was almost always very close to 2. Therefore, the Ni magnetic moment was fixed at 2.

## 3. Results and discussion:

### 3.1. Validation of the potential

Overall, 1400 training data for FLiBe, 866 training data for FLiBeCr, 739 training data for FLiBeNi, and 1339 training data for FLiBeCrNi were generated during the active sampling and used for training the ACE potential. Table 1. shows the root mean squared errors (RMSE) of the training and testing sets for each fitted potential. The training and testing errors are very close, demonstrating that the fitted potentials are not overfit. Due to the diversity of the training data selected by the active learning procedure, the error of the testing data is actually slightly less than the training data. The parity plots of the errors are provided in the supplementary data. It can be observed that the force errors of different atoms are not in the same range. For the main atoms of the salt (Li, Be, F) the errors are smaller than 100 meV/Å. Based on the previous works in the literature [2,3] such accuracy is sufficient to reproduce the structural and thermophysical properties of the salt. For Ni and especially Cr the force errors are higher than the main atoms of the salt by about one order of magnitude. These larger errors are not particularly surprising as a recent article [33] has discussed the challenges of modeling d-block elements, especially early transition metals, with MLIPs due to the complexities in their interatomic interactions. However, it is important to make sure, as best we can, that these larger errors for Cr and Ni do not lead to errors in the modeling of the properties of interest.

To assess the performance of the potential for Cr and Ni we make three comparisons to the AIMD that include both structural and kinetic properties, namely, the radial distribution functions (RDF), angular distribution functions (ADF), and diffusion coefficients. First, we compare the RDF and ADF of Cr-F and Ni-F bonds between AIMD and ACE-MD. For each single impurity system, we ran one AIMD and one ACE-MD simulation for 20 ps in NVT ensemble at $T$ = 973 K. Figure 3 shows the RDFs. Both curves show good agreement between AIMD and ACE. For Cr-F the average bond lengths calculated by DFT and ACE are 1.98 Å and 1.99 Å and the coordination numbers up to the first nearest neighbor are 5.30 and 5.52, respectively. For Ni-F the average bond lengths calculated by DFT and ACE are 1.99 Å and 1.97 Å and the coordination numbers up to the first nearest neighbor are 5.56 and 5.41, respectively. Figure 4 shows the ADFs. Both curves show good agreement between AIMD and ACE. It should be noted that due to the existence of only one Cr or Ni in each system, the RDF and ADF curves are not as smooth as the curves that one might obtain for the other cations in the system and to get smoother curves and subsequently better bond length and coordination number calculation, the length of the simulations should be much longer to get better statistics. Here, we only aimed to show the general agreement between the curves and did not run long AIMD simulations, which are computationally expensive.

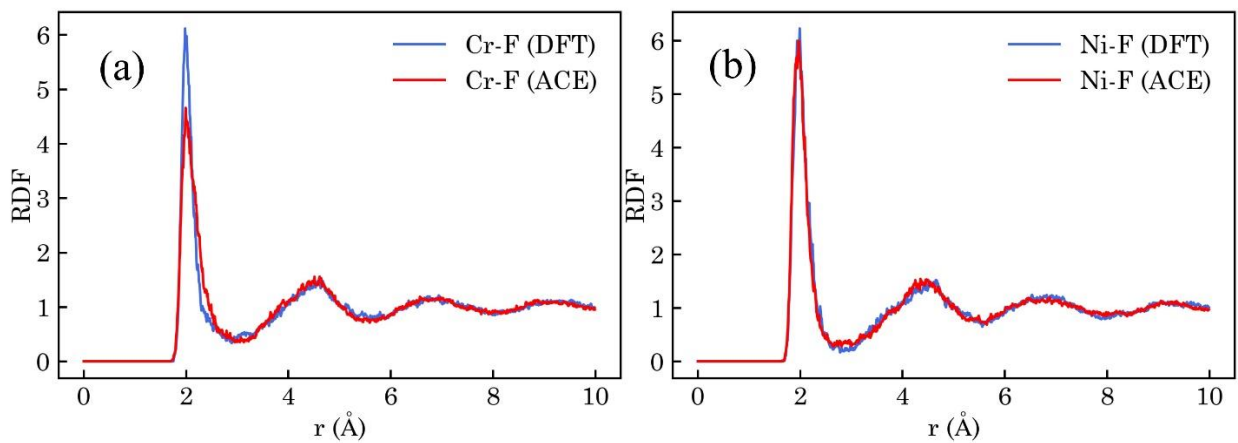

**Figure 3**. *Comparison of radial distribution functions (RDFs) of (a) Cr-F and (b) Ni-F between DFT and ACE.*

**Table 1.** *Root mean squared errors (RMSE) of training and testing sets. The values are in the units of meV/atom for energies, and meV/Å for forces.*

|  | Training | Testing |
|---|---|---|
| Energy (meV/atom) | 1.5 | 1.3 |
| Force (meV/Å) |  |  |
| Li | 43 | 41 |
| Be | 50 | 45 |
| F | 49 | 42 |
| Cr | 317 | 296 |
| Ni | 122 | 120 |

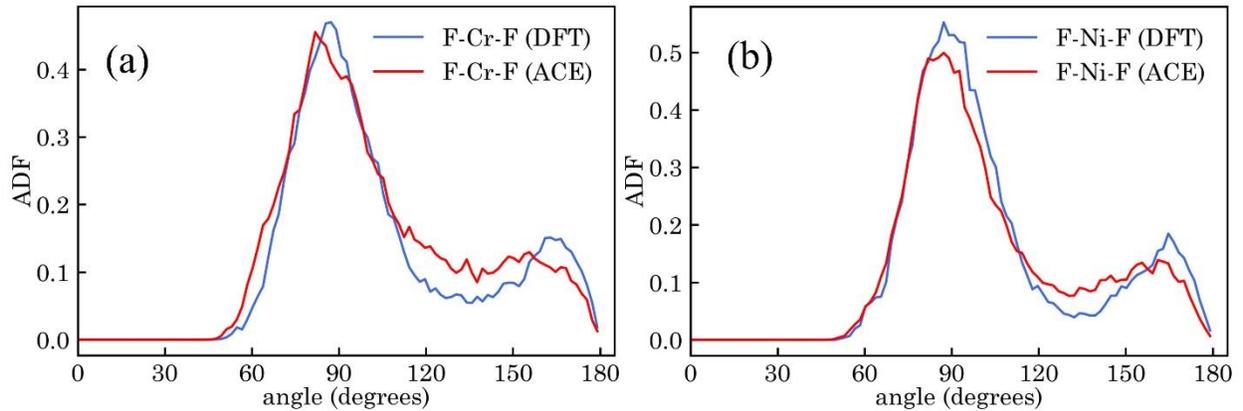

**Figure 4**. *Comparison of angular distribution functions (ADFs) of (a) F-Cr-F and (b) F-Ni-F between DFT and ACE.*

We then compared the diffusivities predicted by AIMD and ACE MD by calculating the slope of the mean squared displacement (MSD) using Einstein's relation [34]

$$D = \frac{1}{6} \lim_{t \to \infty} \frac{d}{dt} \left[ \frac{1}{N} \frac{1}{n_t} \sum_{i=1}^{N} \sum_{j=1}^{n_t} \left( r_i(t_j + dt) - r_i(t_j) \right)^2 \right] \qquad Eq. 8$$

where $D$ is the self-diffusion coefficient, $N$ is the number of atoms, $n_t$ is the number of time origins, and $r_i$ is the position of atom $i$. For a fair comparison of diffusivity between AIMD and ACE, one would need to run very long simulations with a big cell containing many atoms to get an accurate estimate of the diffusivity from each simulation. This is not practical with AIMD as we are limited to a small simulation cell with short simulation time. In addition, the presence of only one Cr and Ni in the systems makes the diffusivity calculation very difficult to converge robustly even in ACE-MD, and particularly in AIMD. Given these limitations, for a reasonable comparison, we ran one AIMD and several ACE-MDs and examined whether the diffusivity from AIMD falls within the range of diffusivities obtained from ACE-MD. We ran one AIMD simulation for 20 ps, and 100 separate MD simulations with ACE for 100 ps all in NVT ensemble at $T$ = 973 K, and compared the range of values we obtained for diffusivities of various elements. The simulations were performed for FLiBeCr and FLiBeNi systems (see Fig 1.) and the results are shown in Figure 5. In Figure 5 the blue dots are the diffusion coefficients for each element calculated from each of the 100 different ACE-MD simulations and the red star is the diffusivity calculated from AIMD simulations. It can be seen that the calculated diffusivities for Ni and Cr from AIMD are within the predicted diffusivities from ACE-MD and the spread of the diffusivities predicted for Ni and Cr are similar to Li, although Li has lower force errors and a higher number of atoms within the simulation cell.

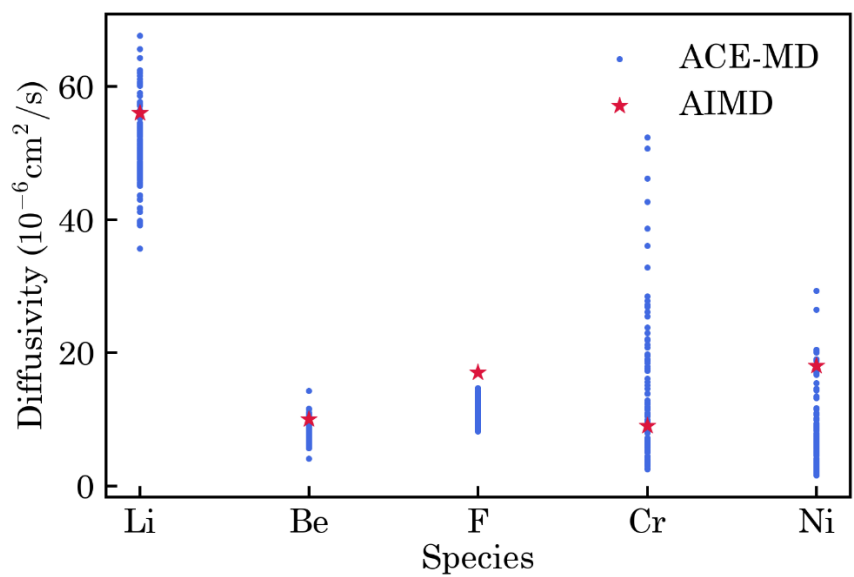

***Figure 5***. *Diffusivities predicted by ACE-MD (blue dots) and AIMD (red star).*

Figure 6 compares the predicted diffusivities of Cr and Ni using AIMD with the histogram of the predictions using ACE-MD. The diffusivity values are non-negative and do not follow a normal distribution. However their logarithm seems to be approximately normally distributed and we used the logarithm of diffusivities to obtain Gaussian models assessing means and errors. For each histogram of logarithms of diffusivities, we fitted a Gaussian distribution, and the mean and the standard deviations are shown in Figure 6. For Cr, the AIMD-predicted diffusivity falls within one standard deviation of the mean of ACE-MD predictions, and for Ni, the AIMD-predicted diffusivity falls within two standard deviations of the mean of ACE-MD predictions. In both cases, it is highly plausible that AIMD and ACE-MD values are drawn from the same distribution. Based on this analysis we conclude that the kinetics of Cr and Ni in FLiBe are being calculated with enough accuracy for exploring their behavior, and that this accuracy is comparable to that of the other three elements (Li, Be, F) with lower force errors.

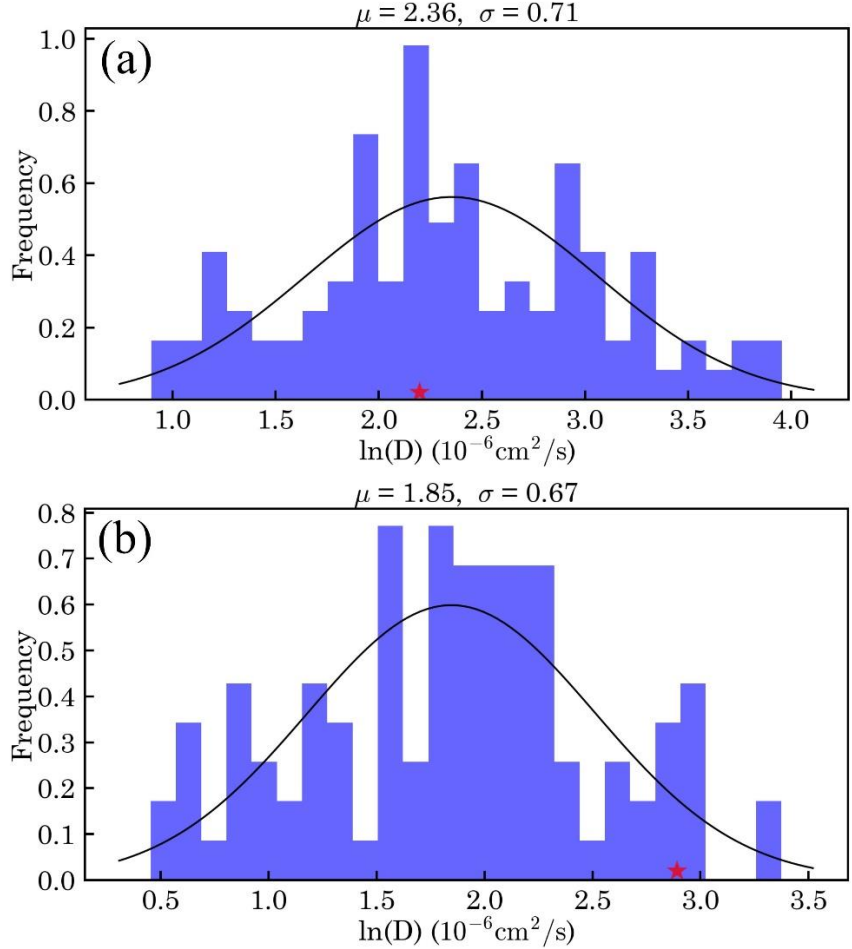

***Figure 6***. *Comparison of the predicted diffusivities of (a) Cr and (b) Ni using AIMD (red star)*

*and ACE-MD (histogram). The values are on logarithmic scale.*

### 3.2. Complexation of Cr-Ni in FliBe

Figure 7 shows the time evolution of g(r) between Cr and Ni for the three systems. At each simulation length, the g(r) is the average of g(r) from the beginning up to that length. For the FLiBeCrNi_small system, after 60 ns of the simulation, the g(r) converges and the curves at 60 ns up to 200 ns essentially overlap. For the FLiBeCrNi_medium system, the g(r) also converges after 60 ns, and for For FLiBeCrNi_large system, the convergence happens somewhere between 60 ns and 80 ns. We take the final g(r) of each simulation (at 200 ns) and compare them in Figure 8(a).

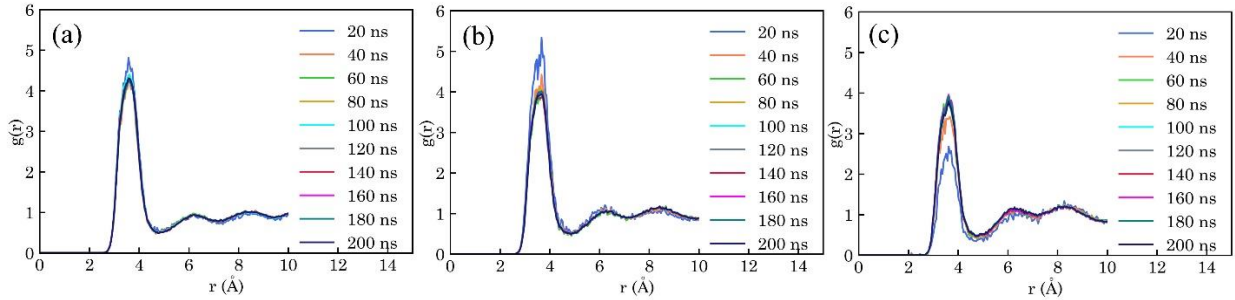

***Figure 7***. *Comparison of the g(r) between Cr and Ni of (a) small, (b) medium, and (c) large system at every 20 ns of the simulations.*

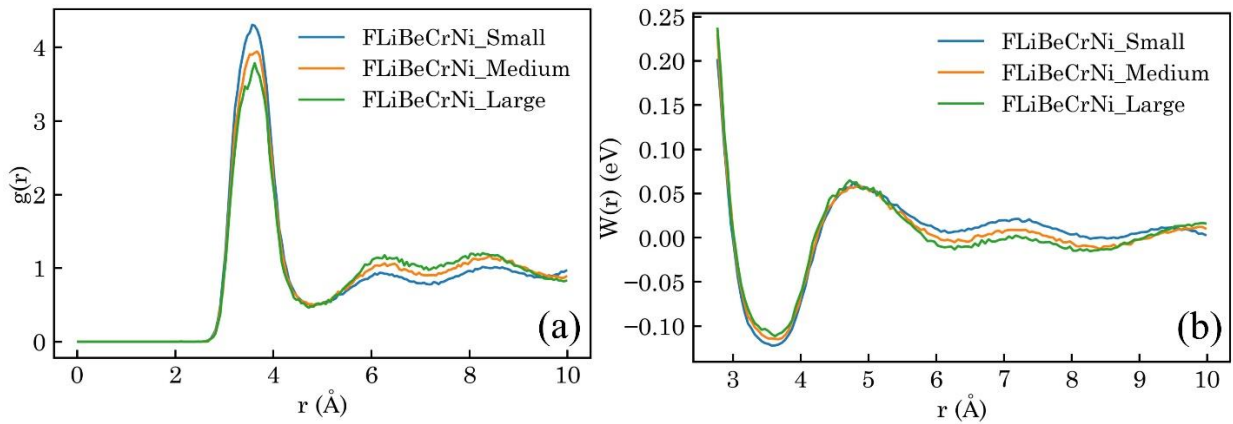

***Figure 8***. *Comparison of the g(r) (a) and PMF (b) between Cr and Ni of each simulated system after 200 ns of simulation.*

The average lengths of the simulation boxes after 200 ns of simulation are 13.5 Å, 17.0 Å, and 19.5 Å respectively. All three systems show a g(r) peak at about 3.6 Å which is the second coordination shell of Cr (or Ni) with one F ion between Cr and Ni (see Fig. 10). There is a noticeable decrease in the height of g(r) between the small and the medium systems, but the change in g(r) from medium to large system is margial. Simulating an even larger system would have required a much longer time to converge g(r) and the obtained curve would likely be very similar to the g(r) of the FLiBeCrNi_large system, so we did not pursue simulating a larger system. We believe that the g(r) from FLiBeCrNi_large system represents an adequately converged curve with respect to size for calculating PMF. Figure 8(b) compares the PMF curves obtained from each simulation using Equation 1 which are essentially Gibbs free energy profiles of the system as a function of the distance between Cr and Ni. Here due to the logarithmic nature of the relationship between g(r) and W(r), the difference between the free energy curves is very slight, and the minimum values for the curves are -0.122 eV, -0.115 eV and -0.112 eV for small, medium and large systems respectively. As at larger distances the g(r) approaches 1, the W(r) approaches zero, and for the FLiBeCrNi_large system the binding Gibbs free energy $\Delta G_b = -0.1116 \pm 0.0005\ eV$ between when the impurities are farther apart and when they are in second coordination shell of each other. The reported significant figures and associated error of 5 x 10$^{-4}$ is obtained by dividing the 200 ns simulation in three equal sections (about 66 ns each) and calculating the error in the mean of the $\Delta G_b$ from each section. The RDF and PMF curves of these sections are provided in the supplementary data. We note that this error is purely that associated with the numerical uncertainty from the molecular dynamics simulation and is provided as a guide to numerical convergence. This error does not reflect the errors compared to the real system associated with other factors, for example, the DFT and potential fitting. These latter errors are almost certainly orders of magnitude larger than this numerical value.

This result implies that the system's free energy is lower when Cr and Ni are in each other's vicinity compared to when they are isolated. As a consequence of this result, we expect that Cr and Ni will enhance each other's solubility. To explore this aspect further we imagine a scenario where some Cr has already been dissolved (since it is more reactive than Ni) and consider the impact on the solubility of Ni in the presence of this Cr. Based on values provided in Ref. [8] extracted from HSC Chemistry v. 9.0 [35], the redox potentials of Cr/CrF$_2$ and Ni/NiF$_2$ versus

F$_2$/F$^-$ in FLiBe at $T$ = 973 K are $V_{Cr}$ = -3.93 and $V_{Ni}$ = -3.06 volts respectively. Using equations 9 to 11 we calculated the molar concentration of unbound or isolated Cr ([Cr$_{ub}$] as modeled in the FLiBeCr system), unbound Ni ([Ni$_{ub}$] as modeled in the FLiBeNi system), and bound Ni ([Ni$_b$]; Ni that is within the same simulation cell as Cr as modeled in FLiBeCrNi_large system)

$$[Cr_{ub}] = \exp\left(\frac{n_e(V - V_{Cr})}{k_B T}\right) \qquad\qquad Eq.\,9$$

$$[Ni_{ub}] = \exp\left(\frac{n_e(V - V_{Ni})}{k_B T}\right) \qquad\qquad Eq.\,10$$

$$[Ni_b] = [Ni_{ub}] \times [Cr_{ub}] \times K_{eq} = [Ni_{ub}] \times [Cr_{ub}] \times \exp\left(\frac{-\Delta G_b}{k_B T}\right) \qquad Eq.\,11$$

In Equation 9 (and correspondingly in Equation 10) $V$ is the salt potential referenced to F$_2$/F$^-$, $V_{Cr}$ is the standard redox potential of chromium, $k_B$ is the Boltzmann constant, $T$ is temperature and $n_e$ is the number of electrons transferred in the redox reaction Cr + 2F $\rightarrow$ CrF$_2$   ($n_e$ = 2 in both equations 9 and 10 as both chromium and nickel are in the oxidation states of 2). For example, if $V = V_{Cr}$ then [Cr$_{ub}$] = 1 mol/l and the higher the salt potential, the higher the solubility of Cr in the salt. Equation 11 is based on the binding free energy of Equation 2 and the assumption of ideal configurational entropies for the Ni and Cr. Equation 11 also considers anything not in the nearest cation neighbor shell to be unbound. Equation 11 is used to calculate the molar concentration of bound Ni, which also represents the amount of extra Ni that is dissolved in the salt due to the presence of Cr. Figure 9 shows the variation of the dissolved unbound Cr and Ni, bound Ni, and the total (bound + unbound) dissolved Ni with respect to the salt potential F$_2$/F$^-$. At lower salt potentials Ni is predominantly dissolved by the potential of the salt itself. By increasing the salt potential, the concentration of the dissolved Cr ([Cr$_{ub}$]) in the salt increases, and consequently the concentration of the dissolved Ni due to the presence of Cr ($[Ni_b]$) increases. At around $V$ = -3.99 volts where the blue and the green lines intersect, the amounts of dissolved unbound and bound Ni are equal, and above that potential Cr complexation effectively draws more Ni to the salt than the salt potential itself. As mentioned in Section 2.1, our calculations are based on the interaction of a single Cr and a single Ni while in the real systems, there may be multiple Cr and Ni atoms in a region. Our calculation assumes that Cr-Cr and Ni-Ni and multiple Ni with Cr or multiple Cr with Ni interactions are negligible. Although these results are approximate, they are useful in providing a qualitative guide of the likely scale of the effects.

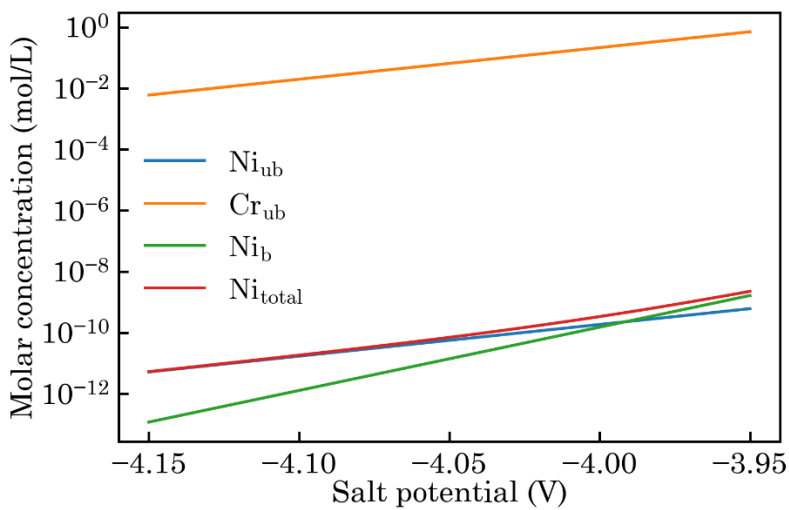

***Figure 9***. *The molar concentration of unbound or isolated Cr ($Cr_{ub}$), unbound, bound, and total Ni ($Ni_{ub}$, $Ni_b$, $Ni_{total}$) with respect to the salt potential referenced to $F_2/F^-$.*

We also studied the structural features of the FLiBeCrNi_large system by calculating the Warren-Cowley parameters for the constituents of the salt around chromium. In the first coordination shell of Cr (see Fig. 10), one naturally finds fluorine ions as they have ionic bonding. In the second coordination shell, there should mostly be the other three cations that are attached to the chromium via fluorine ions. The Warren-Cowley short-range order parameter captures the deviation from the random order of neighboring atoms around a central atom and is calculated as follows

$$\alpha_{ij} = 1 - \frac{Z_{ij}}{\chi_j . Z_i} \qquad\qquad Eq.\,12$$

where $\alpha_{ij}$ is the Warren-Cowley parameter, $Z_{ij}$ is the partial coordination number of atom $j$ around the central atom $i$, $Z_i$ is the total coordination number around atom $i$, and $\chi_j$ is the concentration of atom $j$ in the simulation cell. If $\alpha_{ij}=0$ there is no short-range ordering around the central atom $i$ meaning the atoms are randomly dispersed. $\alpha_{ij} < 0\ (> 0)$ means that atoms of $j$ type tend to segregate to (away from) atom $i$. We calculated $\alpha_{ij}$ only for the second coordination shell around Cr and only considered Be, Li, and Ni atoms to calculate the total coordination number and atomic concentration in the simulation cell. The calculations are based on the MD simulation of FLiBeCrNi_large system for 200 ns in NPT ensemble at $P$ = 0 bar and $T$ = 973 K. We obtained $\alpha = -0.13$ for Li, $\alpha = 0.26$ for Be, and $\alpha = -0.79$ for Ni. The coordination numbers are calculated from the shaded areas under the partial RDF curves as depicted in Figure 11.

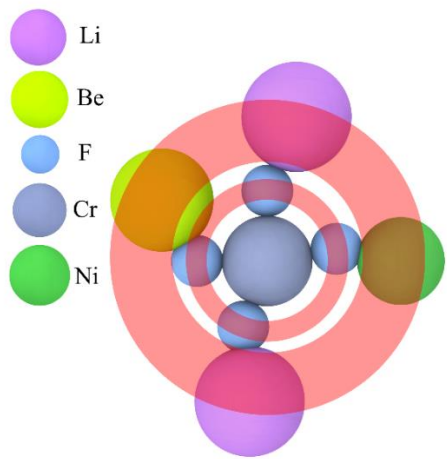

***Figure 10****. Illustration of the first and the second coordination shell around Cr ion.*

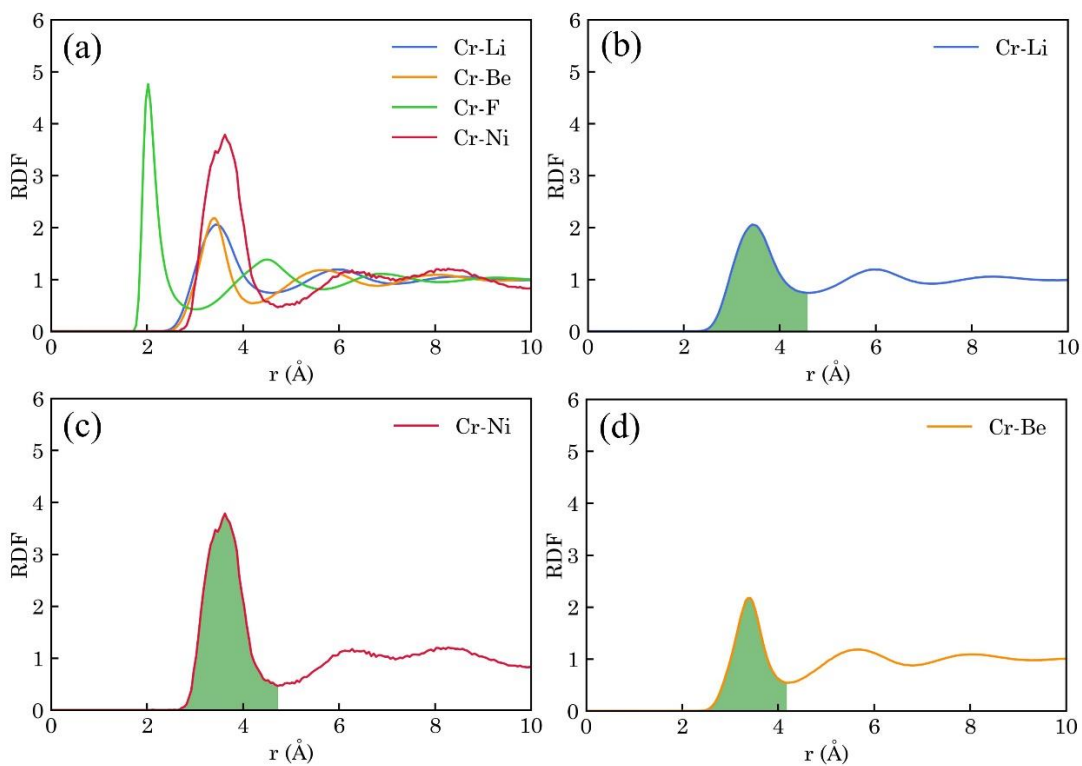

***Figure 11****. (a) Partial radial distribution function (RDF) around atom Cr. The shaded areas in panels (b), (c), and (d) are used to calculate the coordination numbers of atoms Li, Be, and Ni, respectively.*

Between the three calculated parameters, the value of $\alpha$ is negative but close to zero for Li and positive but close to zero Be, implying they are close to randomly dispersed around Cr during the simulation with Be weakly being pushed away from Cr and Li weakly attracted to Cr. For Ni, the value is a larger negative number than for Li, which suggests that the Ni ion was significantly overrepresented in the second coordination shell of Cr compared to random arrangements. In other words, Ni tended to maintain a closer distance to Cr than expected by random chance and therefore can be considered to have formed a complex of some kind with Cr. As previously mentioned the average size of the cell for this simulation was 19.5 Å which allows a maximum distance between Cr and Ni of ~9.75 Å parallel to a cell edge and ~16.87 Å on the diagonal line, considering periodic boundary conditions. From Figure 11 (d), it can be seen that after the first peak (when Ni is one fluorine away from Cr), there are three other peaks at around 6, 8, and 10 Å where Ni ions are two, three, and four fluorines away from Cr, which leaves Ni ion multiple neighbor shells for separating from Cr. Therefore, we do not think that the closeness of Ni and Cr was somehow forced by the size of the simulation cell and rather it has naturally occurred due to the energetics of the system.

It is interesting as a check on the consistency of our approaches to compare the MD measured Warren-Cowly parameter to that we can estimate from simple thermodynamics using the binding free energy between Ni and Cr. Using equation 11 and $[Ni_{ub}] = [Cr_{ub}] = 0.229 \ mol.L^{-1}$ (1 atom in a cell with the volume of 7238 Å³), one can calculate $[Ni_b] = 0.119 \ mol.L^{-1}$. This means that there should be $\frac{[Ni_b]}{[Ni_{ub}]} = 52\%$ more Ni in the second coordination shell (SCS) of Cr, compared to the case where the are no interaction between Ni and Cr.

If there were no interaction between Cr and Ni, and Ni was homogenously dispersed across the simulation cell one could calculate an average of $\frac{N_{Ni} \times V_{SCS}}{V_{total}} = \frac{1 \times 366}{7238} = 0.05$ Ni atoms in the SCS of Cr, where $N_{Ni}$ is the total number of Ni atoms in the simulation cell, $V_{SCS}$ is the volume of SCS of Cr and $V_{total}$ is the total volume of the simulation cell. Knowing that there must be 52% more atoms of Ni in the SCS one can predict a partial coordination number of 0.05*1.52 = 0.076 for Ni in SCS of Cr and accordingly a Warren-Cowly parameter of $\alpha = -0.56$, which is close to -0.79 calculated using our previous method. The discrepancies are likely due to simplifying assumptions in the binding energy model, which treats the sytems as having just two unique states (bound and unbound), and the assumption of homogeneity, which is not exactly correct since the FLiBeCrNi system has multiple coordination shells around Cr, with some having more cations and some more anions.

To study the effects of the complexation of Cr and Ni on the kinetics of these ions, we calculated the diffusivities of unbound and bound Cr and Ni by running 100 MD simulations of FLiBeCr, FLiBeNi, and FLiBeCrNi_small systems in NVT ensemble at $T$ = 973 K, similarly to what was discussed in Section 2.1. The results are shown in Figure 12. The error bars represent the error in the mean. For chromium, we observed a marginal increase in the diffusivity going from an unbound system to a bound system (FLiBeCrNi_small system), but the difference is within the errors of the calculation. For nickel, there is an apparently statistically significant increase in the calculated diffusivity from the unbound to the bound system, suggesting the Ni diffusion may be slightly increased by its coupling to Cr. However, the effect is very small (about a 25% increase). Overall, due to the very slight difference between the diffusivities of bound and unbound chromium and nickel, we do not predict a significant change in the diffusivity of Cr or Ni when complexed.

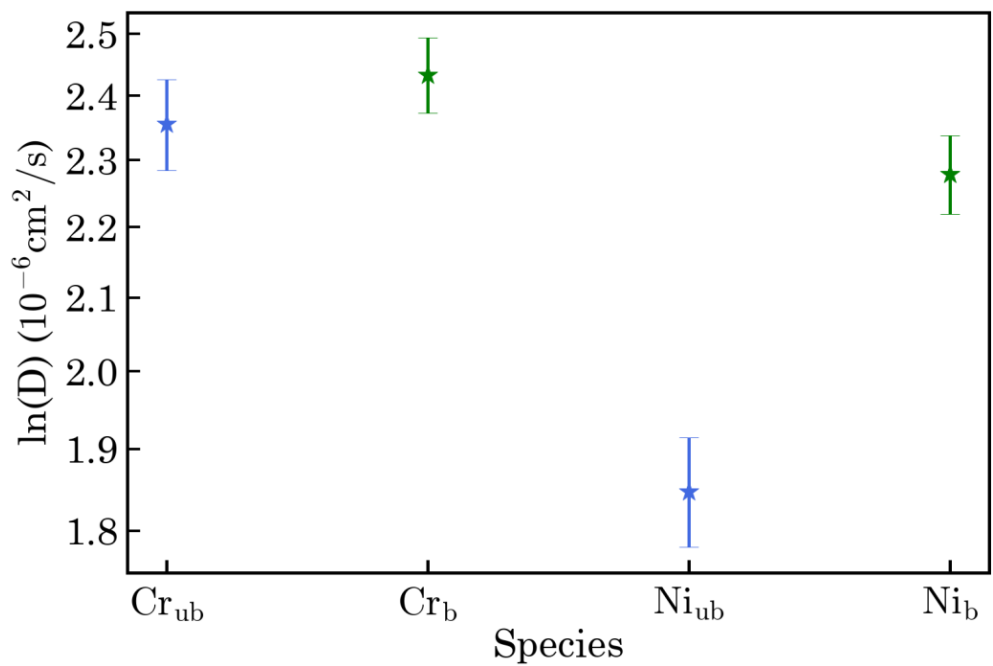

*Figure 12*. *Diffusivity of unbound and bound chromium (Cr$_{ub}$ and Cr$_b$) and unbound and bound nickel (Ni$_{ub}$ and Ni$_b$) at 973 K. The results are shown in logarithmic scale so the extent of the error bars would appear equal on both sides.*

## 4. Conclusion

In this work, we presented an application of MLIPs in studying the effects of impurity complexation on corrosion of structural materials in molten salts, which was neither practical by AIMD and likely would be challenging to do accurately with existing classical potential MD simulation. We discussed the challenges in developing MLIPs for molten salts with impurities and provided a methodology to overcome the challenges by using an active learning scheme based on the D-optimality criterion for data generation. We then developed ACE potentials, ran long MD simulations, calculated the PMF between Cr and Ni, and studied the structure, and kinetics of the impurities in the salt. Our results show that there is a weak binding between Cr and Ni with a reaction free energy of $\Delta G = -0.112$. This binding, although weak, may impact some properties of the salt, for example using this binding energy we showed that the dissolution of Ni increases in the presence of Cr. Structural studies of FLiBe containing both impurities suggest that Cr and Ni mostly reside in each other's second nearest neighbor shells connected by one F ion. Kinetic studies suggest a very weak effect of this interaction on diffusion coefficients, with perhaps a slight increase (~25%) in the diffusivity of a Ni when in the presence of a Cr compared to when it is isolated. It should be noted that we have simplified the problem by considering 1 Ni and 1 Cr, but in general, it is possible to use the same approach presented in this study to model multiple impurities of the same or different kinds and investigate the effect of impurities on the thermophysical and structural properties of molten salts. Such a modeling requires developing potentials for the whole composition of FLiBe + $CrF_2$ + $NiF_2$ which is outside the scope of this work and is left for future works.

## Acknowledgment

We gratefully acknowledge support from the Department of Energy (DOE) Office of Nuclear Energy's (NE) Nuclear Energy University Programs (NEUP) under award # 21-24582.

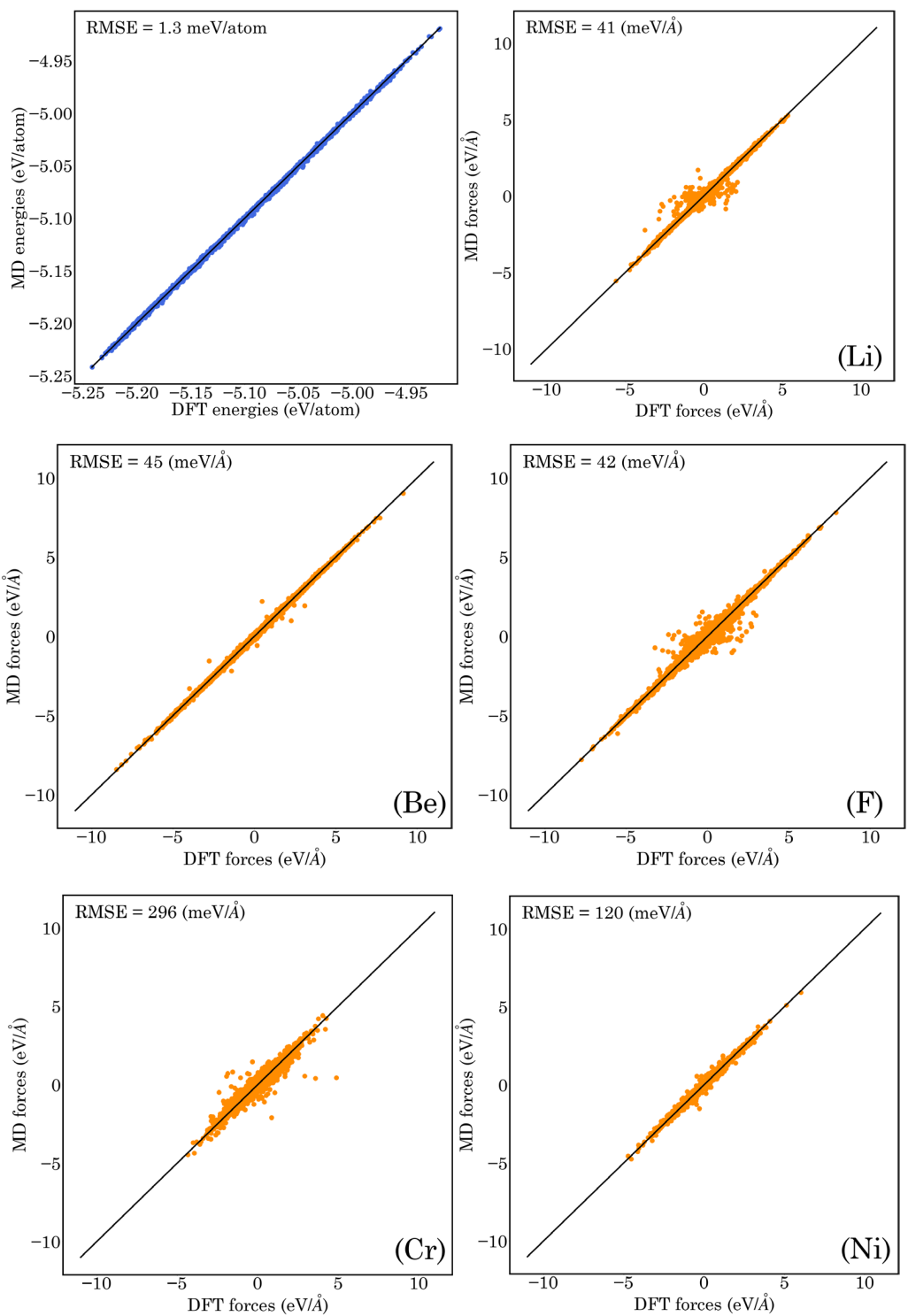

***Figure S1***. *Parity plots of the testing set for energies (top left) and forces of each atoms type (the rest of the panels. The atoms type is indicated in each panel).*

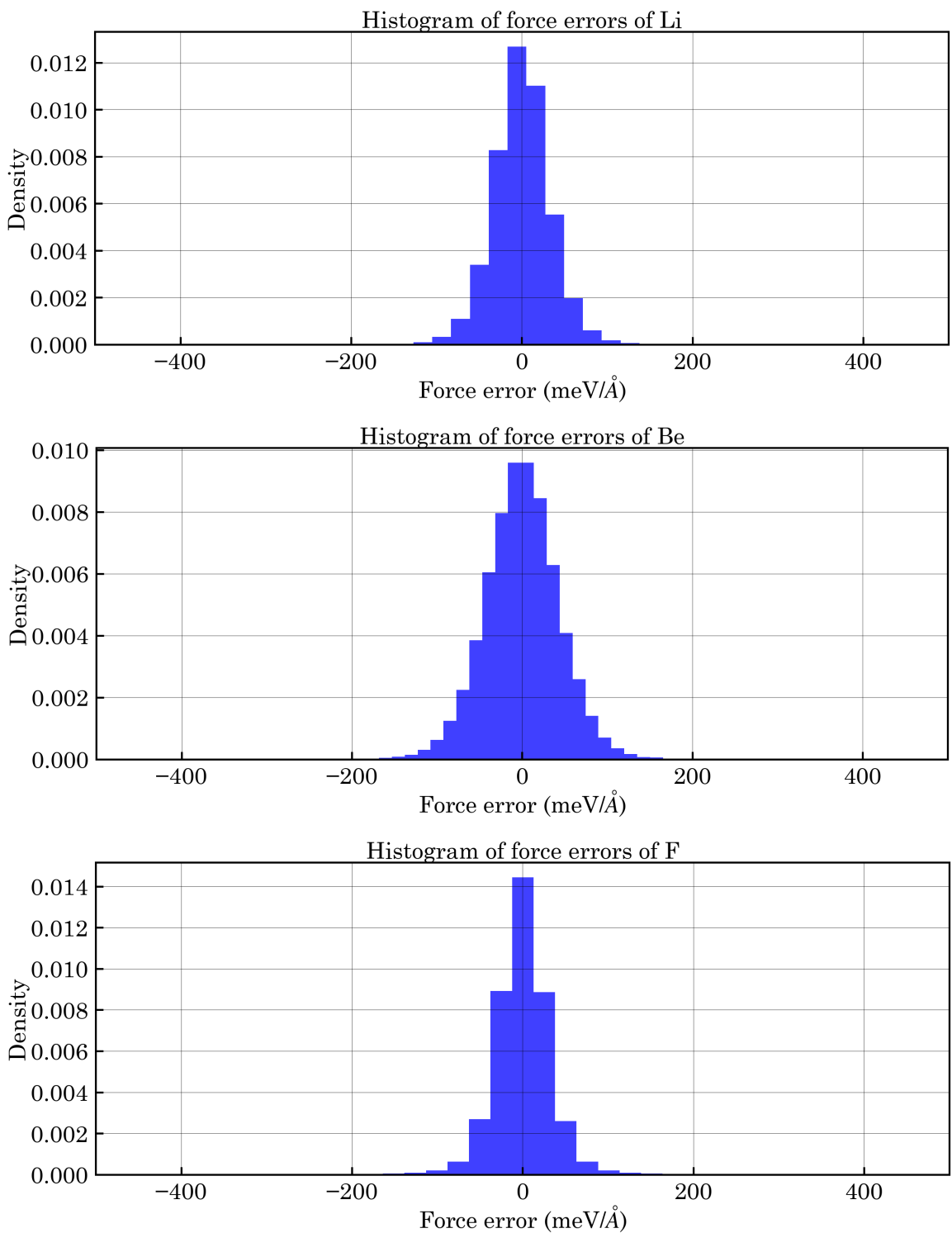

***Figure S2****. Histogram of the force errors for Li (top), Be (middle), and F (bottom).*

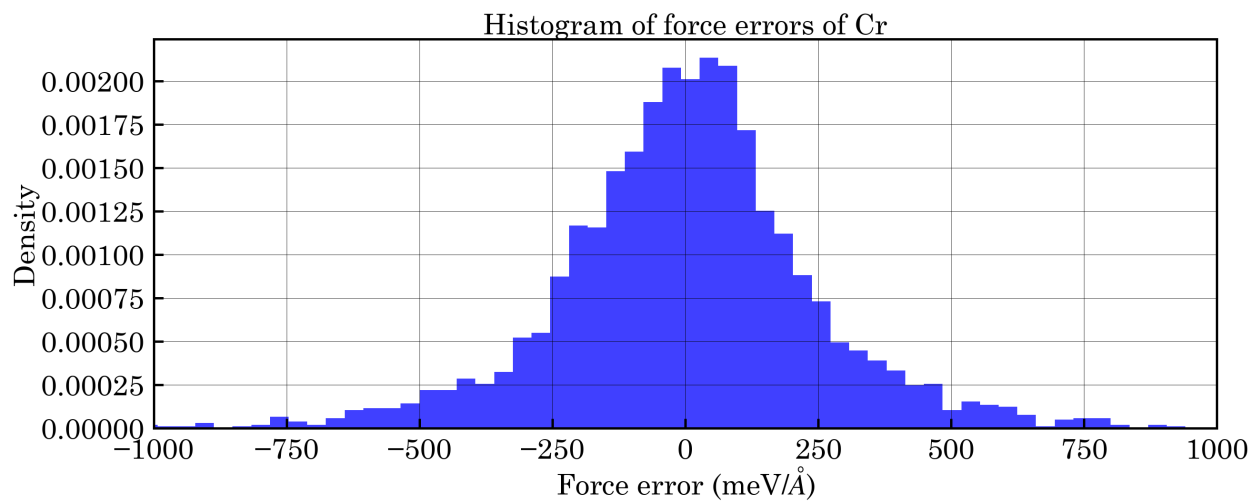

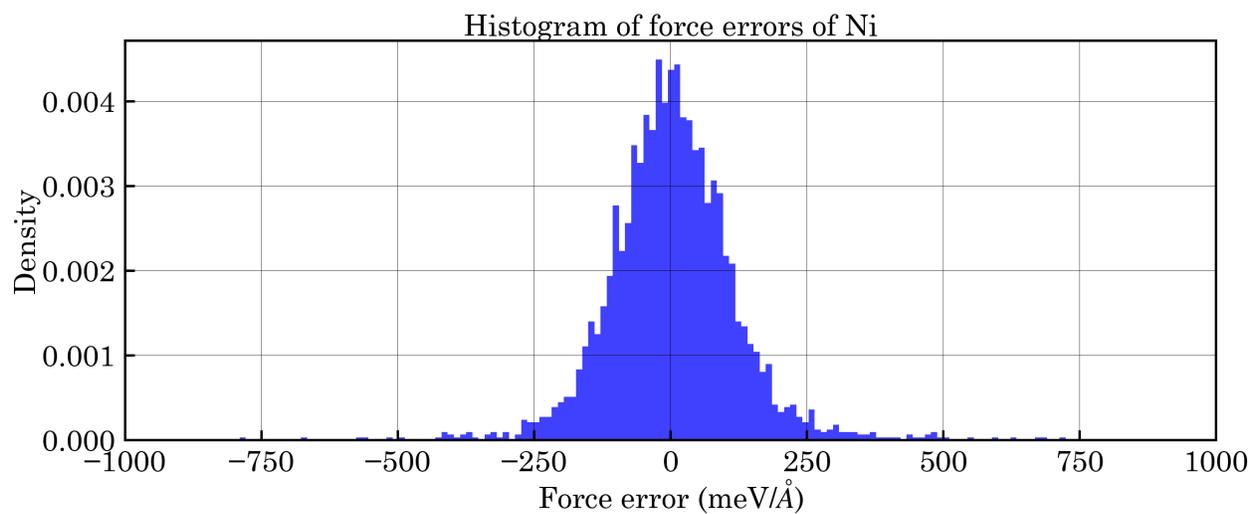

*Figure S3*. *Histogram of the force errors for Cr (top, and Ni (bottom).*

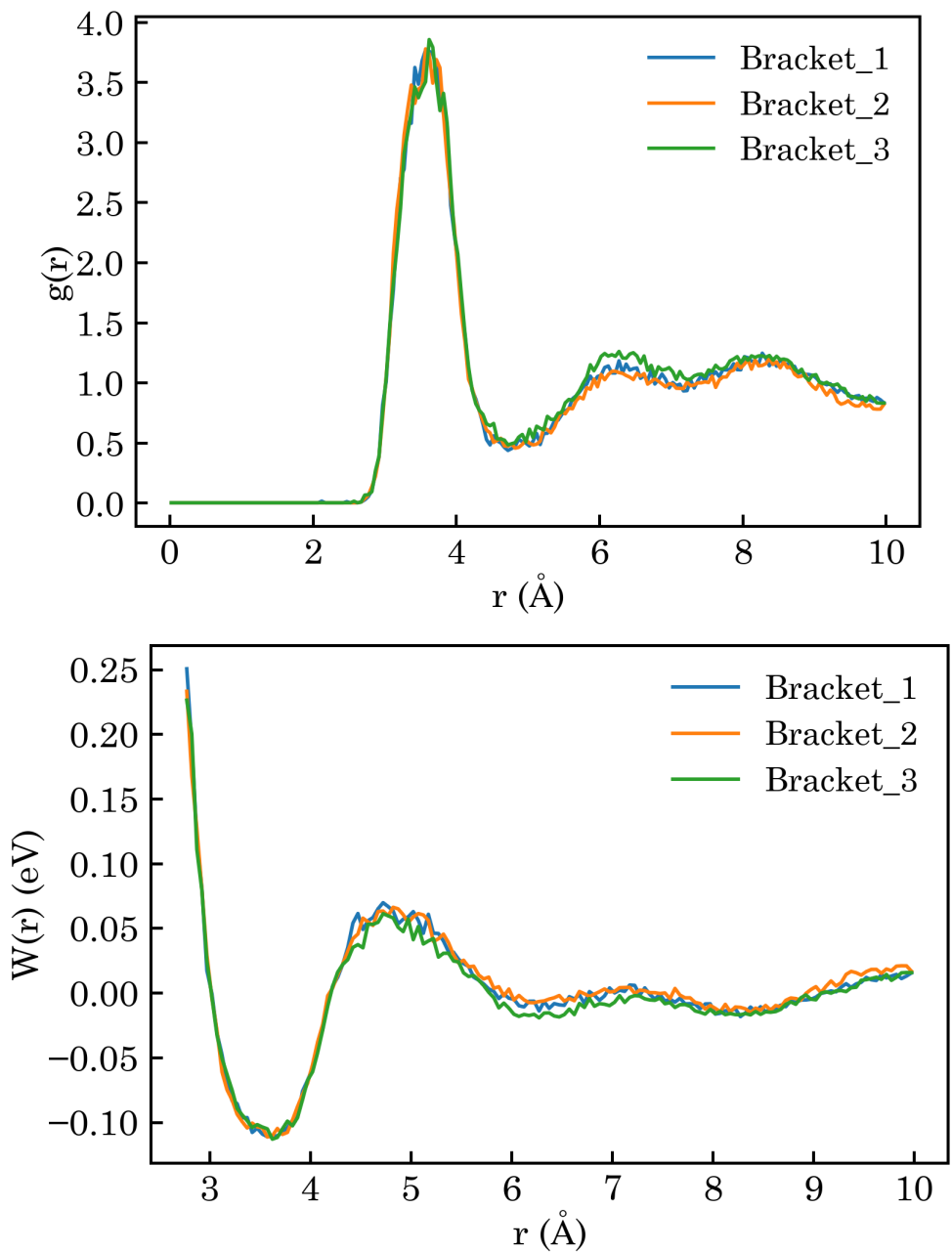

***Figure S4**. Comparison of the RDF (top) and PMF (bottom) between three sections of the 200 ns simulation of FLiBeCrNi_large system. Each section or bracket is about 66 ns of the total simulation.*